# The Spectrum of Interstitial Solute Energies in Polycrystals

Malik Wagih and Christopher A. Schuh[1]

*Department of Materials Science and Engineering, Massachusetts Institute of Technology,*
*77 Massachusetts Avenue, Cambridge, MA, 02139, USA*

**Abstract**

A computational method is presented to measure the spectrum of segregation energies for an interstitial solute at grain boundaries (GBs) in a polycrystal. For the Pd-H system, that spectrum of GB interstitial segregation energies is found to consist of a mixture of two normal-like distributions, characteristic of GB octahedral and tetrahedral sites. We also derive a thermodynamic isotherm to calculate the equilibrium segregation state in a polycrystal that recognizes the need for referencing two interstitial site types, and show that it conforms well to experimental data in the Pd-H system.

The segregation of solute atoms at grain boundaries (GBs) in polycrystalline materials can affect a wide range of structural and functional properties [1–4]. The grain boundary network, due to its structural disorder, can provide a multitude of local atomic environments (or site-types) that are more favorable to solute atoms than the ordered bulk lattice environment. We have recently examined in Refs. [5–8] the nature and frequency of different site-types at the GB in polycrystals, and how this translates into a wide spectrum of segregation energies for solute atoms; this contradicts the classical wisdom of McLean-type models that assume a single site-type at the GBs, i.e., a single averaged segregation energy. Previous work in this vein, however, has been limited to substitutional solute elements, and as yet the site spectrum for interstitial solutes is a purely speculative matter [9–11].

In this letter, we develop a modeling approach to measure, for the first time, the more complex interstitial solute segregation spectrum. More specifically, we aim to develop a methodology that can address four main questions:

1. What characteristics identify interstitial sites in polycrystals, especially at the GBs?
2. What are the possible energetic (enthalpic) states for an interstitial solute in the bulk (intra-grain) region and at the GBs?
3. What is the appropriate thermodynamic isotherm to solve for the equilibrium segregation state, while capturing the spectral nature of interstitial segregation?
4. What functional form can be used to quantify interstitial segregation in a binary system based on a few simple parameters?

We address these questions for a model system for which there is both prior experimental data [9,10] as well as a viable interatomic potential [12,13], namely, the segregation of a H solute atom in a Pd base metal. This model system is used to establish a framework that should be easily extensible to other binary interstitial alloys.

The first challenge is to identify interstitial sites in a polycrystal. A perfect fcc unit cell has well defined interstitial sites, i.e., tetrahedral and octahedral sites, but it is not as straightforward to define such sites in polycrystals, especially for the disordered regions at the GBs. We propose that Voronoi tessellations may be used as a method of determining the location of interstitial sites within any system of atoms; the approach has precedent in simpler and more ordered systems [14–17], and here we apply it for full polycrystals with large samples of highly disordered boundaries. In this approach, Voronoi polyhedra [18] are constructed around the atoms in the system, and the Voronoi vertices are used as coordinates for interstitial sites, from which relaxation can yield ideal solute locations; Voronoi vertices within 1.0 Å of one another are merged into a single point. In Fig. 1(a) and (b), we demonstrate the validity of this approach by applying it to the perfect Pd fcc supercell; the method correctly identifies all octahedral and tetrahedral sites in the fcc cell (1 octahedral and 2 tetrahedral sites per atom in an fcc cell).

---

[1] Corresponding author.
 *Email address:* schuh@mit.edu (Christopher A. Schuh)

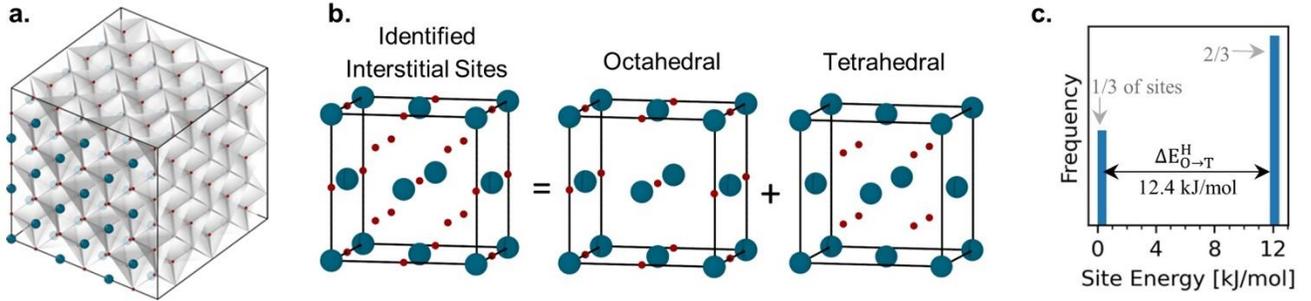

Fig. 1: (a) A visualization of the construction of Voronoi polyhedra to identify interstitial sites in an fcc supercell. (b) For a perfect fcc Pd cell, the Voronoi-based method correctly identifies all interstitial sites, which include (as illustrated by hand) one octahedral and two tetrahedral sites (small red spheres) per Pd atom (large teal spheres). (c) The energetics of a H atom occupying the identified interstitial sites in a perfect Pd cell, using the octahedral site – the most favorable bulk site – as a reference.

Fig. 2(a) shows a 20x20x20 nm³ Pd polycrystal built with Atomsk [19] using Voronoi tessellations, and subsequently annealed at 600 K under a Nose-Hoover thermostat/barostat for 500 ps using a time step of 1 fs, to slightly grow the grains and relax the GBs into a more realistic configuration [20]. After cooling at 3 K/ps to 0 K, we relax the polycrystal using the conjugate gradient minimization algorithm, followed by a force minimization using the Fire algorithm [21] (with a force convergence criterion of $10^{-10}$ eV/ Å). The annealed sample has 545,794 atoms, distributed among 16 grains with an average grain size of ~10 nm. The adaptive-common neighbor analysis method [22] is used to identify the Pd GB sites, where we assume all non-fcc solvent atoms are GB atoms. This results in a total of 97,765 GB solvent atoms and a GB atomic site fraction of 17.9%. All atomistic simulations are performed using LAMMPS [23,24], and atomic structures are identified and visualized using OVITO [25].

Using the Voronoi-based site identification method, we identify ~1.6 million interstitial sites in the (20 nm)³ Pd polycrystal, which corresponds to a ratio of ~2.9 interstitial sites per Pd atom, which indicates that the GB region has slightly fewer interstitial sites than expected for ideal fcc crystal structures (with an interstitial-to-solvent ratio of three-to-one; cf. Fig. 1(b)). (We note that the ratio of interstitial-to-solvent atoms is controlled by how aggressively we merge nearby interstitial sites at the GB, i.e., a shorter merging cutoff distance yields more interstitial sites, and vice versa). Next, we need to determine which interstitial sites belong to the GB region. If an interstitial site is surrounded by GB atoms, we can easily count it as a GB interstitial site, but it is not immediately clear what neighbor fraction of GB solvent atoms define a GB solute, especially given the long-range stress fields of interstitial solutes, which may profitably interact with GBs at a significant distance. The H solute site energy (or segregation energy in case of a GB interstitial site) is defined as $\Delta E_i^{seg} = E_{i=sol}^{poly} - E_{c=sol}^{poly}$ where $E_{i=sol}^{poly}, E_{c=sol}^{poly}$ are the total relaxed energies of the polycrystal with a solute atom occupying an interstitial site (i), versus the energetically favorable bulk interstitial site (which, in the case of Pd(H), is the octahedral site), respectively. Both states are relaxed using the Fire algorithm [21] with a force convergence criterion of $10^{-3}$ eV/ Å. We chose an octahedral site in the polycrystal that is surrounded by 4 nm of fcc Pd atoms in all directions (i.e., the center of an 8 nm sphere) as the reference site for $E_{c=sol}^{poly}$ to avoid long-range interactions with the GB. We note that, in this notation, a negative $\Delta E_i^{seg}$ signifies an interstitial site that is favorable to solute segregation, i.e., more energetically favorable than the bulk interstitial (octahedral) sites.

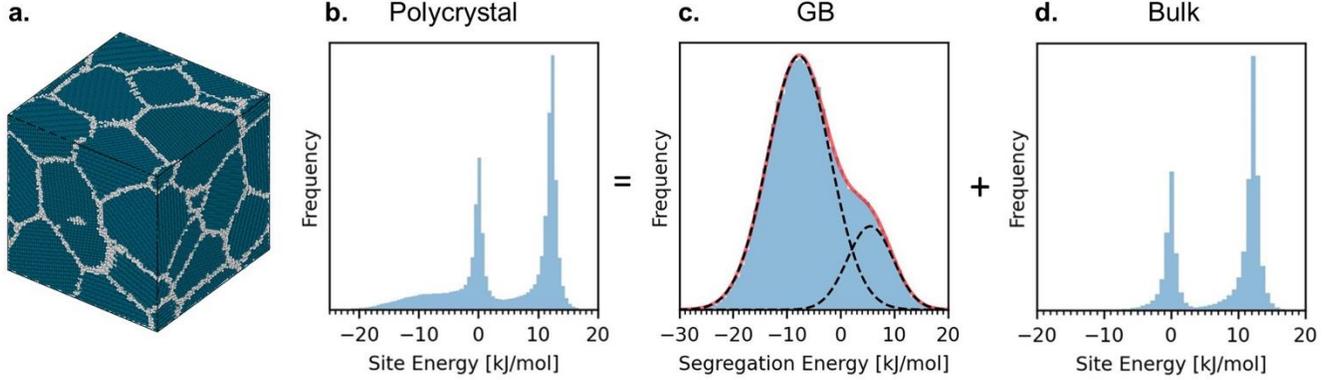

Fig. 2: (a) The annealed 20x20x20 nm$^3$ Pd polycrystal with 16 grains of ~10 nm size; GB solvent (Pd) atoms are highlighted in white as identified by the adaptive common neighbor analysis method. (b) The spectrum of H energetics for all ~1.6 million interstitial sites in the polycrystal (using the bulk octahedral site as a reference), which can be separated into (c) a spectrum for GB interstitial segregation energies, which is captured by a mixture (solid red line) of two Gaussians (black dashed lines); and (d) a spectrum for bulk energies that separates into two sharp energy levels representing the bulk octahedral ($\Delta E_i^{seg} = 0$ kJ/mol) and tetrahedral sites ($\Delta E_i^{seg} = $ ~12 kJ/mol).

The full spectrum of such H interstitial energies in the Pd polycrystal is presented in Fig. 2(b), and exhibits multiple peaks. Which of these many energies belong to GB sites is addressed further in Fig. 3(a), where we show the variation of the H insertion energy as a function of their minimum distances from the nearest Pd GB atom. Very close to the GB, up to a distance of 2.7 Å (which is the nearest neighbor distance for fcc Pd), there is a significant variation in the energetics of the interstitial sites (spanning from -25 to 15 kJ/mol). Farther than that, the energies sharply separate into two horizontal energy levels centered at ~0 kJ/mol and ~12 kJ/mol. These two levels represent the bulk octahedral and tetrahedral interstitial sites, respectively, which we confirm by making the calculation on the ideal fcc supercell from Fig. 1, with the result shown in Fig. 1(c). The rapid bifurcation and sharp narrowing of the two peaks with distance indicate that the bulk interstitial sites are not essentially affected by long-range interactions with the GB and maintain their separate character with a characteristic ~12 kJ/mol energy difference between them. Therefore, it is reasonable to define GB interstitial sites as all sites within a cutoff distance of 2.7 Å from Pd GB sites, which yields a site fraction of 25.3 % of interstitial sites in the present polycrystal. As illustrated in Fig. 3(b), which shows only interstitial sites in the polycrystal without Pd atoms, the GB interstitial network (yellow sites) is similar to that of pure Pd (Fig. 2(a)), but somewhat thicker because interstices that border GB Pd atoms are considered GB solute sites.

With this definition of GB sites, in Fig. 2(c) and 2(d) we now can separate the spectra of segregation and site energies for H atoms in the GBs and bulk, respectively. The bulk spectrum in Fig. 2(d) is a reasonable match to the idealized two-spike bulk distribution in Fig. 1(c), while the GB spectrum in Fig. 2(c) is, we believe, the first such computed spectrum for interstitial GB segregation. Interestingly, the GB spectrum exhibits two bumps, which are well-characterized as a mixture of two Gaussian distributions at the GB. The normal-like nature of the GB spectra is in line with earlier observations for substitutional GB segregation [5,7,8], with the difference being that substitutional GB segregation is characterized by a single Gaussian-like distribution, while interstitial GB segregation is characterized by a mixture of two Gaussian-like distributions. Intriguingly, the centers (means) of the two Gaussians differ by 12 kJ/mol, the same energy difference between octahedral and tetrahedral sites in the bulk, suggesting that even in the more disordered GB environment it is possible to differentiate semi-octahedral and semi-tetrahedral sites.

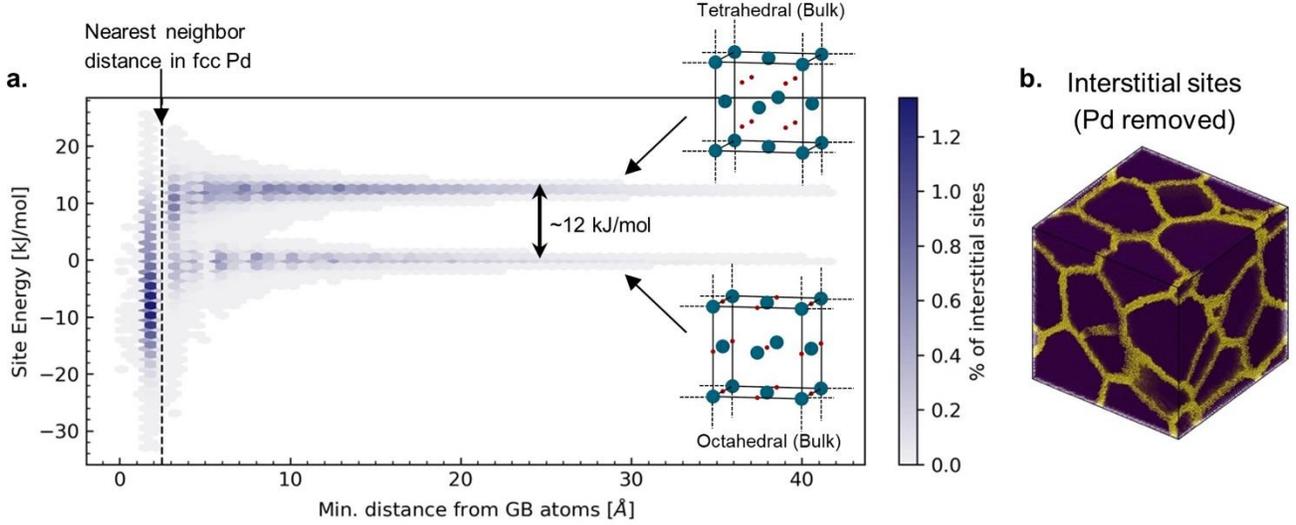

Fig. 3: (a) The variation of H energetics for the occupation of interstitial sites as a function of the minimum distance from a Pd GB atom. (b) A visualization of all interstitial sites in the Pd polycrystal with Pd atoms removed; we define the GB interstitial sites (yellow sites) as all sites within a distance of 2.7 Å (nearest neighbor distance in Pd) from a Pd GB atom.

The simple form of the interstitial site energy spectra in Fig. 2 suggests that we may reduce the ~1.6 million energies that went into its measurement into just a few parameters. For the bulk, we propose simplifying its discrete distribution, Fig. 2(d), into two site-types similar to that of a perfect fcc cell as shown in Fig. 1(c). The first is a reference octahedral site-type with $\Delta E_i^{seg} = 0$ and a bulk site fraction of 1/3, and the second is a tetrahedral site-type with $\Delta E_i^{seg} = \Delta E_{O \rightarrow T}^H = 12.4$ kJ/mol and a bulk site fraction of 2/3. For the GB, we follow our suggestion above of a mixture of two Gaussians, which reduces the distribution function $F_i^{gb}(\Delta E_i^{seg})$ to six parameters:

$$F_i^{gb} = w_1 \cdot \frac{1}{\sqrt{2\pi}\,\sigma_1} \cdot \exp\left[-\frac{1}{2}\left(\frac{\Delta E_i^{seg} - \mu_1}{\sigma_1}\right)^2\right] + w_2 \cdot \frac{1}{\sqrt{2\pi}\,\sigma_2} \cdot \exp\left[-\frac{1}{2}\left(\frac{\Delta E_i^{seg} - \mu_2}{\sigma_2}\right)^2\right] \quad (1)$$

where w is the weight, μ is the mean, and σ is the standard deviation of a given Gaussian distribution. In Table 1, we list the values of all seven parameters required to characterize interstitial segregation in the Pd(H) system – $\Delta E_{O \rightarrow T}^H$, $w_1$, $\mu_1$, $\sigma_1$, $w_2$, $\mu_2$, and $\sigma_2$.

Table 1: The required 7 parameters to characterize interstitial GB segregation for Pd(H) with Eq. (4).

| | |
|---|---|
| $\Delta E_{O \rightarrow T}^H$ | 12.4 kJ/mol |
| $w_1$ | 0.2 |
| $\mu_1$ | 5.5 kJ/mol |
| $\sigma_1$ | 4.3 kJ/mol |
| $w_2$ | 0.8 |
| $\mu_2$ | -7.8 kJ/mol |
| $\sigma_2$ | 5.8 kJ/mol |

To predict the equilibrium segregation state, we derive a spectral thermodynamic isotherm for the interstitial problem. In a polycrystal, the total interstitial solute concentration $X^{tot}$ is fixed, and is divided among bulk and GB sites by their atomic site fractions, $f^c$ and $f^{gb}$, respectively [26]:

$$X^{tot} = f^c X^c + f^{gb} \bar{X}^{gb} \qquad (2)$$

where $X^c, \bar{X}^{gb}$ are the equilibrium solute concentration at the bulk and GB, respectively; $f^{gb}+f^c = 1$, and $f^{gb} = 25.3\%$ as computed earlier for this Pd polycrystal. We define the concentrations as fractions of occupied interstitial sites; this can be easily converted to H/Pd concentration by dividing by 3 (the ratio of interstitial sites to Pd atoms). The interstitial bulk concentration ($X^c$) is divided among octahedral ($X_O^c$) and tetrahedral ($X_T^c$) sites according to their site fractions $X^c = \frac{1}{3}X_O^c + \frac{2}{3}X_T^c$. We can re-express Eq. (2):

$$X^{tot} = (1 - f^{gb}) \cdot \left[\frac{1}{3}X_O^c + \frac{2}{3}X_T^c\right] + f^{gb}\bar{X}^{gb} \qquad (3)$$

As $X_O^c$ is the concentration for the bulk reference site (i.e., $\Delta E_i^{seg} = 0$, and the energetics of other sites-types are measured in reference to it), Eq. (3) can be expanded as [5,10,27,28]:

$$X^{tot} = (1 - f^{gb}) \cdot \left[\frac{1}{3}X_O^c + \frac{2}{3} \cdot \left[1 + \frac{1-X^c}{X^c}\exp\left(\frac{\Delta E_{O \to T}^H}{kT}\right)\right]\right] + f^{gb} \cdot \int_{-\infty}^{\infty} F_i^{gb} \cdot \left[1 + \frac{1-X^c}{X^c}\exp\left(\frac{\Delta E_i^{seg}}{kT}\right)\right]^{-1} d\Delta E_i^{seg} \qquad (4)$$

To validate the spectral isotherm of Eq. (4), we performed Monte Carlo simulations, where solute atoms are randomly added to the polycrystal, and then are allowed to swap interstitial sites using the Metropolis criterion, with an acceptance probability of $\min[1, \exp(-\Delta E^{swap}/kT)]$, with $\Delta E^{swap} = \Delta E_2^{seg} - \Delta E_1^{seg}$, where the subscripts 1, 2 refer to interstitial sites before and after swapping, respectively. The simulations are run until equilibration for a total of 1000 swaps per solute atom. In Fig. 4(a) and (b), we compare the results of the Monte Carlo simulations (scatter points) against that of the isotherm (solid lines), Eq. (4); it is clear that the isotherm captures the correct physics of segregation in the system, but at a much lower computational cost, using only 7 parameters (Table 1) in a straightforward evaluation of Eq. (4), instead of the full discrete ~1.6 million energies and millions of computations used in the Monte Carlo simulations.

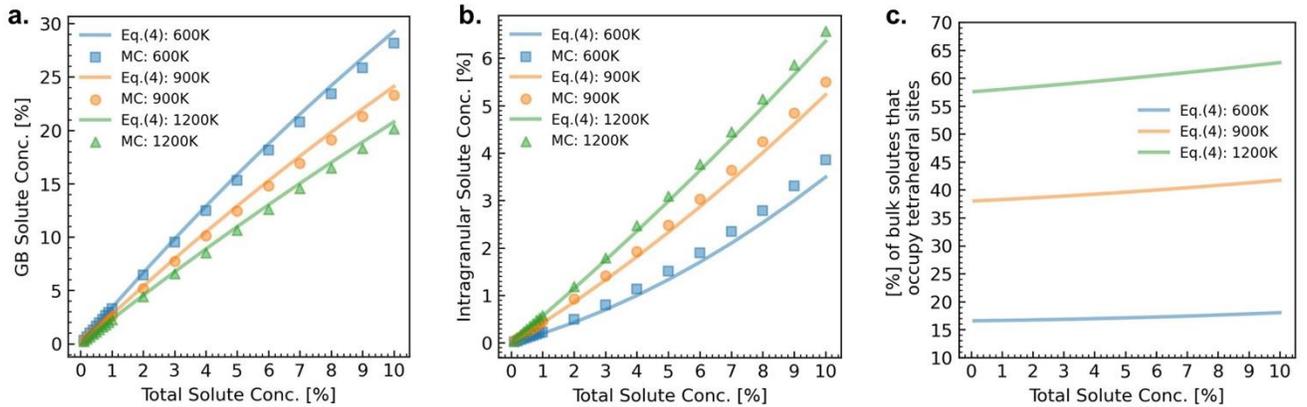

Fig. 4: A comparison of the results of equilibrated Monte Carlo (MC) simulations that use the full ~1.6 million energies against the predictions of Eq. (4) using only the 7 parameters listed in Table (1) for (a) GB solute concentration and (b) Intragranular (bulk) solute concentration at different compositions and temperatures. (c) The fraction of solute atoms in the bulk that occupy the enthalpically unfavorable tetrahedral interstitial sites, as predicted by Eq. (4), for different compositions and temperatures.

One interesting and perhaps unexpected observation from these analyses is that solute atoms will occupy the enthalpically unfavorable bulk tetrahedral sites across the composition-temperature space. Since the tetrahedral sites are twice as many as the octahedral sites, they are entropically more favorable. The literature often overlooks this point by omitting unfavorable bulk sites; for example, in the case of Pd(H), the octahedral sites would be assumed to be the sole sites occupied by solute atoms [9,10]. However, as we illustrate in Fig. 4(c), the fraction of solute atoms in the bulk that occupy the enthalpically unfavorable tetrahedral sites are non-negligible across the composition-temperature space, with the value reaching over 60% of solute atoms in the bulk (intra-grain region).

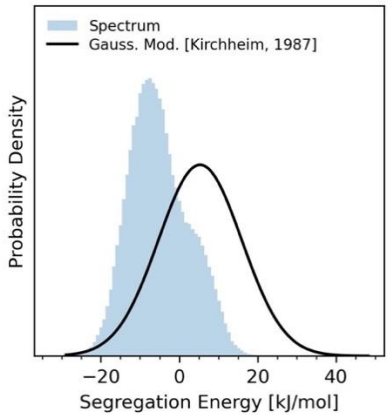

Fig. 5: A comparison of the computed GB spectrum for Pd(H) in this work versus the proposed Gaussian by Kirchheim and co-workers (solid line) based on their interpretation of experimental data.

Finally, we note that the calculated spectrum for hydrogen segregation in Pd is consistent with observations made by Kirchheim and co-workers [9,10] in the 1980s based on their seminal experimental work on charging nanocrystalline Pd with hydrogen. Kirchheim proposed that H segregation at the GB can be explained by a distribution of site types at the GB, and derived an isotherm similar to Eq. (4) using Fermi-Dirac statistics. There are two major differences between Kirchheim's analysis and our current results. First, Kirchheim assumed a single Gaussian distribution of segregation energies or site types at the GB, as shown in Fig. 5, but here we demonstrate there is a mixture of two Gaussians (or normal-like distributions). Second, Kirchheim ignored tetrahedral sites in the bulk because of their enthalpic unfavorability; however, in this study, we demonstrate that entropic contributions will result in a non-negligible fraction of solute atoms occupying these sites. The forward use of computed spectra such as those in Fig. 2 to predict experimental outcomes, and the inverse use of experiments to extract better forms of that spectrum, should both be more plausible in light of the present observations, and we leave such connections to future work.

In conclusion, we have presented a computational methodology for measuring the spectrum of segregation energies for interstitial solutes segregating to GBs. We have outlined the proper thermodynamic isotherm to calculate the equilibrium solute content at the GB, noting that the new observations made in this work suggest that one must take into account the multiplicity of interstitial site-types within both the bulk and the GBs for a rigorous approach to this problem. Only a few parameters are needed to capture the complex energetics of interstitial segregation in binary alloys (we used 7 in the present example of Pd(H): six parameters to characterize the mixture of two Gaussians at the GB—a mixture of semi-tetrahedral and semi-octahedral sites—and one to describe the energy difference between the bulk tetrahedral and bulk octahedral sites). We hope this letter inspires the next necessary steps to catalog such parameters and create comprehensive databases for interstitial segregation across the alloy space.

## Acknowledgments

This work was supported by the US Department of Energy, Office of Basic Energy Sciences under grant DE-SC0020180.